# Applications of the Lorentz-Abraham-Dirac equation in long-term dynamics

Marijan Ribarič and Luka Šušteršič

Jožef Stefan Institute, p.p. 3000, 1001 Ljubljana, Slovenia
E-mail: marjan.ribaric@masicom.net

ABSTRACT
   **Motivation**: The Lorentz-Abraham-Dirac equation is appropriate for?
   **Objective:** To elucidate the problems.
   **Method:** Perturbation modeling of the long-term dynamics, whereby the Lorentz-Abraham-Dirac equation is interpreted as the lowest order, asymptotic differential relation for the velocity of the radiating point-like mass, and higher order terms are added for more precise modeling.
   **Results:** In addition to mass and charge, supplementary kinetic constants of accelerated charged particles are defined.
   **Application:** Experimental study of the kinetic properties of charged particles by the improved model of the effects of the radiation reaction force.

## Content

1. Introduction
2. The basic conceptual question of classical electrodynamics
3. A four-momentum balance equation for an electrified point-like mass
   *3.1. A differential four-momentum balance equation*
   *3.2. Comments*
4. Newtonian equation of motion for a charged particle
5. Modification of the Lorentz-Abraham-Dirac equation
   *5.1. An asymptotic, relativistic differential relation for EPM´s velocity*
   *5.2. Applications to calculations of the external forces and trajectories*
6. Discussion
   *6.1. Experimental assessment of RRF's dynamic effects*
   *6.2. Testing theories*
7. Conclusions
8. References

## 1. Introduction

In classical electrodynamics we study the implications of Lorentz forces between electric charges and currents: and these forces affect their sources. About twenty years ago we published a monograph [1] about certain open questions of classical electrodynamics. Here we reconsider the basic open question about the Newtonian equations of motion in classical electrodynamics, since it turned out we should have followed the advice of English philosopher G.E. Moore in 1903, cf. [2, Ch.1], which we rephrase as

➢ **Preposition about questions:** To avoid difficulties and disagreements one should not attempt to answer questions, without trying discovering precisely what question it is which you desire to answer.

To this end we give a breakdown of this open question, itemizing the closely related but distinct subjects we are going to address:

a) We denote by "Newtonian equation for a point like mass" an equation that generalizes Newton's second law by exactly specifying the acceleration by a possibly nonlinear transform of the external force.
b) In Section 2, we point out the basic conceptual question: Can the present classical electrodynamics provide appropriate Newtonian equations of motion that take a *non-iterative* account of the loss of four-momentum by the electromagnetic radiation that results in the *radiation reaction force* (RRF).
c) In Sections 3, to get the simplest example of this question we specify such *differential four-momentum balance equations* for an *electrified point-like mass* (EPM) that would take account of the RRF – to get them is the problem of mathematical physics.
d) In Section 4, we mention the quest for a Newtonian equation of motion for a *classical charged particle* – a century old, open question of theoretical physics.
e) In Section 5, in the case of a small and slowly changing external force, we provide a partial answer to the above basic question by reinterpreting the Lorentz-Abraham-Dirac equation, as an asymptotic differential relation for EPM´s velocity. Thereby we introduce new, supplementary kinetic constants of an EPM, which we name after Dirac, Bhabha and Eliezer, which introduced them.
f) In Section 6, we propose a particular *experimental assessment* of these kinetic constants, which take account of RRF's dynamic effects.

g) We consider in [13], how one can model the dynamics of point-like objects by differential equations that are not Newtonian equations of motion.

## 2. The basic conceptual question of classical electrodynamics

Around 1602 Galileo began classical mechanics with the study of pendulums by an innovative *combination of experiment and mathematics*. Thereby he clearly stated that the laws of nature are mathematical. However, if we only slightly rub a pendulum and electrify it, classical electrodynamics provided so far no non-iterative, Newtonian model of its swinging motion, which effects the loss of energy by electromagnetic radiation. We feel that a detailed experimental observation of such a swinging motion would be welcome as physics is based on measurements.

According to Jackson [3], the basic trouble with the classical electrodynamics is that we are able to obtain and study relevant solutions of its basic equations only in two limiting cases: "… one in which the sources of charges and currents are specified and the resulting electromagnetic fields are calculated, and the other in which external electromagnetic fields are specified and the motion of charged particles or currents is calculated… . Occasionally… the two problems are combined. But the treatment is a stepwise one -- first the motion of the charged particle in the external field is determined, neglecting the emission of radiation; then the radiation is calculated from the trajectory as a given source distribution. It is evident that this manner of handling problems in electrodynamics can be of only approximate validity." Thus within the framework of classical electrodynamics there is no non-iterative, Newtonian modeling of interaction between electromagnetic fields and their sources, which we can use for consideration or prediction of experimental data. Consequently, we have only a partial physical understanding of such classical electromechanical systems where we cannot neglect this interaction.

In what follows we intend to broaden our understanding by considering an electrified point-like mass, which seems to provide the simplest case of such an interaction.

### 3. A four-momentum balance equation for an electrified point-like mass

In classical mechanics, the simple, idealized Newtonian model of a point-like mass is highly instructive and widely applicable, e.g. for studying the trajectories of planets. So it seems like it might be useful to have such a Newtonian model about the motion of an electrified point-like mass that takes account of RRF.

*3.1. A differential four-momentum balance equation*

Inspired by the point-like mass model, let us try to get better understanding of RRF's dynamic effects by considering the four-momentum of an accelerated EPM. Let the charge $q$ and mass $m$ of this EPM be located around the point $\boldsymbol{r}(t)$, and moving with velocity $\boldsymbol{v}(t)$ under the influence of a mechanical, Lorentz and/or gravitational external force $\boldsymbol{F}_{\text{ext}}(t)$; and let the relation between the four-force

$$f(t) = \gamma(\boldsymbol{\beta} \cdot \boldsymbol{F}_{\text{ext}}, \boldsymbol{F}_{\text{ext}}), \quad \text{where}$$
$$\gamma(t) = 1/\sqrt{1-|\boldsymbol{\beta}|^2} \quad \text{with} \quad \boldsymbol{\beta}(t) = \boldsymbol{v}/c, \tag{1}$$

and the EPM's four-velocity $\beta(t) = (\gamma, \gamma\boldsymbol{\beta})$ be invariant under Poincaré transformations.[1]

Were $q = 0$, in classical mechanics we would model such a relation by the differential four–momentum balance equation

$$mc\beta^{(1)} = f \quad \text{with} \quad \beta^{(n)} \equiv (\gamma d/dt)^n \beta, \tag{2}$$

which is a heavily used, relativistic Newtonian equation for a point-like mass.

To take account of the RRF, we assume that EPM's electromagnetic radiation is adequately described by the Liénard-Wiechert potentials with singularity at $\boldsymbol{r}(t)$, which emit *at a cyclic motion*, the four-momentum

$$d(\beta^{(1)} \cdot \beta^{(1)})\beta \quad \text{with} \quad d = q^2/6\pi\epsilon_0 c^2, \tag{3}$$

thereby diminishing EPM's four-momentum. Following Schott [4], we find that we must introduce the acceleration four-momentum $B(t)$ *so as* to obtain a complete description of changes to EPM's four-momentum effected by the RRF, cf. 3.2.Comments. Taking these two electromagnetic effects into account, we get from the Newtonian equation of motion (2) the following differential balance equation for EPM´s four-momentum:

$$mc\beta^{(1)} - d(\beta^{(1)} \cdot \beta^{(1)})\beta + B^{(1)} = f. \tag{4}$$

---

[1] We will use the metric with signature $(+ - - -)$, so that $\beta \cdot \beta = 1$.

Dirac [5] concluded that the conservation of four-momentum requires that $B(t)$ is a four-vector valued function of $f$ and $\beta$, and of a finite number of their derivatives, which satisfies the relation

$$\beta \cdot (B + d\beta^{(1)})^{(1)} = 0. \tag{5}$$

Thereafter, Bhabha [6] pointed out that the conservation of angular four-momentum requires that the cross product

$$\beta \wedge (B + d\beta^{(1)}) \tag{6}$$

is a total differential with respect to the proper time of a four-tensor valued function of $f$ and $\beta$, and of a finite number of their derivatives. Assumptions underlying the balance equation (4) are discussed in [1, Chs.9 and 10].

### 3.2. Comments

- The differential four-momentum balance equation (4) and conditions (5) and (6) with $d = 0$ model acceleration of a possibly electrified point-like mass with an internal structure specified by $B(t)$.
- We cannot simplify EPM´s balance equation (4) by disregarding the acceleration four-momentum $B$ carried by EPM´s internal structure; because when $B = 0$ and $d \neq 0$, then the Bhabha condition (6) and the balance equation (4) imply that $f = 0$.
- However, we may just pick

$$B = -d\beta^{(1)} \tag{7}$$

so as to get apparently the simplest possible EPM´s differential four-momentum balance equation:

$$mc\beta^{(1)} - d(1 - \beta\beta\cdot)\beta^{(2)} = f, \tag{8}$$

which is the Lorentz-Abraham-Dirac equation of motion for an electron; the corresponding Lagrangians were recently obtained by Deguchi, Nakano, and Suzuki [20]. The third term of the lhs (8) is known as the Abraham-Lorentz-Dirac RRF.

- Use of equation (8) is physically questionable since
(a) it is not a Newtonian kind of equation of motion, and

(b) it exhibits self-acceleration causing runaway solutions.
In Section 5 we give arguments for using it as an asymptotic differential relation for EPM´s velocity.

- Were certain acceleration four-momentum $B(t)$ given only as a function of EPM´s velocity and of the external force, and satisfy the conditions (5) and (6), the differential four-momentum balance equation (4) would be a relativistic Newtonian equation for the particular EPM specified by this $B(t)$. In [1, Secs.10.1 and 10.2] and [7], we pointed out seventeen qualitative properties that we are expecting from a Newtonian equations of motion for a physically realistic EPM.
- It is still an open question whether we can somehow augment the continuous classical electrodynamics with the concept of a point-like charge, which is presently only a common and handy computational device. We generalized it by an expansion in terms of co-moving moments of time-dependent, moving charges and currents [8].
- EPM seems to be the simplest generalization of classical mechanics concept of a point-like mass to such a mathematical model of classical electrodynamics that takes into account RRF's dynamic effects. We considered it extensively in our monograph [1], naming an EPM a classical point-like charged particle. We found this name to be misleading, since we did not intend it to be synonymous with the concept of an elementary physical particle. So we now rename it "electrified point-like mass".

## 4. Newtonian equation of motion for a charged particle

In 1892, H. A. Lorentz started an ongoing quest to appropriately take account of RRF in modeling the motion of a classical charged particle, cf. [1, 9]. However, there are no pertinent quantitative experimental data. Recently Rohrlich [10] stated that, using physical arguments, he derived from the Lorentz-Abraham-Dirac equation (8) the physically correct Newtonian equation of motion for a classical charged particle:

$$mc\beta^{(1)} = f + (d/mc)(1 - \beta\beta \cdot)f^{(1)}, \qquad (9)$$

provided

$$|(d/mc)(1 - \beta\beta \cdot)f^{(1)}| \ll |f|. \qquad (10)$$

- As pointed out by Rohrlich [11], his equation (9) is a result of a *physical theory*. So it is important to obey its validity limits (10) when testing it by experiments. This equation is not an EPM´s model as specified by the relations (4) to (6).
- Using Langevin equation and the quantum electrodynamics Hamiltonian, G.W. Ford and R.F. O'Connell considered in detail the equation of motion of an electron in papers reviewed by O'Connell [12,]; who comments also on the Rohrlich statements.

### 5. Modification of the Lorentz-Abraham-Dirac equation

We do not know of such an acceleration four-momentum $B(t)$ that results in such a realistic Newtonian equation of motion for an EPM that takes an account of RRF. So it is an open question whether there is one, none or many of them.

Dirac [5], Bhabha [6], and Eliezer [14] pointed out that in addition to $B = -d\beta^{(1)}$, there are also higher order relativistic polynomials in $\beta^{(n)}$ which satisfy both the Dirac and Bhabha conditions (5) and (6). So what would be the physical significance of modified Lorentz-Abraham-Dirac equation (8), were we to assume that EPM´s kinetic properties are modeled by $B(t)$ that is a sum of such polynomials? In 1989, inspired by the expansion of convolution-integrals in terms of derivatives of Dirac's delta function, we proposed that the four-momentum balance equation (4) with a particular combination of such polynomials provides an asymptotic expansion of EPM´s acceleration four-momentum $B(t)$ in the case of a small and slowly changing external force [15].

### 5.1. An asymptotic, relativistic differential relation for EPM´s velocity

We put forward arguments [15; and 1, Chs.9-11] in support of the following hypothesis: Let the acceleration four-momentum $B(t)$ depend in a causal way, just on the values of acceleration $d\boldsymbol{v}(t)/dt$ from the time $t = 0$ on. And let the external force depend on a non-negative parameter $\lambda$ so that $\boldsymbol{F}_{\text{ext}}(t) = \lambda \boldsymbol{F}(\lambda t)$, with $\boldsymbol{F}(t)$ being an analytic function of $t > 0$ and $\boldsymbol{F}(t \leq 0) = 0$. Then, in the asymptote $t \nearrow \infty$, the *n*th derivative $d^n\boldsymbol{v}(t)/dt^n$ of EPM´s velocity is of the order $\lambda^n$ as $\lambda \searrow 0$; and we may approximate the

Application of the Lorentz-Abraham-Dirac equation in long-term dynamics

acceleration four-momentum $B(t)$ up to the order of $\lambda^5$ inclusive so that the EPM´s velocity satisfies in the asymptote $t \nearrow \infty$, up to the order of $\lambda^6$ inclusive, the following differential four-momentum balance equation:

$$mc\beta^{(1)} - d(1 - \beta \cdot \beta)\beta^{(2)}$$
$$+ e_1 [\beta^{(2)} - (\beta \cdot \beta^{(2)} - 1/2\, \beta^{(1)} \cdot \beta^{(1)})\beta]^{(1)}$$
$$+ e_2 [\beta^{(4)} - (\beta \cdot \beta^{(4)} - \beta^{(1)} \cdot \beta^{(3)} + 1/2\, \beta^{(2)} \cdot \beta^{(2)})\beta]^{(1)} \qquad (11)$$
$$+ b_1 \left[(\beta^{(1)} \cdot \beta^{(1)})\beta^{(2)} + 2(\beta^{(1)} \cdot \beta^{(2)})\beta^{(1)} + 7/4\, (\beta^{(1)} \cdot \beta^{(1)})^2 \beta\right]^{(1)} = f,$$

where $d, e_1, e_2$ and $b_1$ are real parameters, cf. [13]. The first term of the relativistic differential relation (11) is due to Einstein; the second one is due to Dirac who calculated that for an electron $d = e^2/6\pi\epsilon_0 c^2$ [5]; the relativistic polynomials in $\beta^{(n)}$ multiplied by $e_1$ and $e_2$ were constructed by Eliezer [14], and that multiplied by $b_1$ is due to Bhabha [6]. So let us refer to $d$ as the Dirac kinetic constant, to $e_1$ and $e_2$ as the Eliezer kinetic constants, and to $b_1$ as the Bhabha kinetic constant.

- As the Lorentz-Abraham-Dirac equation (8) equals the asymptotic differential relation (11) with $e_1 = e_2 = b_1 = 0$, it makes sense to interpret it as the first asymptotic differential relation for EPM´s velocity that that takes some account of the RRF in the case of a small and slowly changing external force. And we can take the differential four-momentum balance equation (11) as an improvement on the Lorentz-Abraham-Dirac equation with regard to the order of parameter $\lambda$ of the external force.

- To get an idea of the significance of the asymptotic differential relation in modeling the motion for a classical charged particle, we consider simple analogous cases in [13].

According to Dirac [5], an electron is such a simple thing that we may assume that $B = -d\beta^{(1)}$. So we put forward

➢ **Hypothesis about the acceleration four-momentum $B(t)$ of an electron.** The Eliezer and Bhabha constants in relation (11) of an electron are negligible.

## 5.2. Applications to calculations of the external forces and trajectories

When the kinetic constants $m, d, e_1, e_2$ and $b_1$ of a particular EPM are known, one may use the differential relation (11), cf. [15] and [1, Sects.11.4 and 11.5], as follows:

**a)** To determine up to the order of $\lambda^6$ inclusive the external force that results in a particular EPM´s trajectory.

**b)** To modify (11) into a variety of approximate Newtonian equations of motion, such as the Rohrlich equation (9), by eliminating through iteration all higher order derivatives of velocity. Then use them to calculate approximate EPM´s trajectories.

**c)** To complete an ansatz, which we use to describe an asymptotic EPM´s trajectory, by expressing its parameters in terms of the external force constants and EPM´s kinetic ones.

**d)** To use it in construction of an empirical formula for the prediction of experimental data about EPM´s trajectories, cf. [16].

## 6. Discussion

### 6.1. Experimental assessment of RRF's dynamic effects

- The Dirac, Eliezer and Bhabha parameters of the asymptotic differential relations (11) are supplementary kinetic constants, additional to the mass of a point-like physical object, possibly electrified. We could assess them by fitting the observed trajectories corresponding to various external forces. To this end we might modify the asymptotic differential relation (11) by replacing the time derivatives with finite difference approximations, choosing step sizes consistent with the experimental data to be fitted. We have no quantitative suggestions about when we may expect to obtain in this way the consistent values of Dirac, Eliezer or Bhabha kinetic constants of an electrified physical object. So let us point out three qualitative conditions:
  a) The object should be point-like in the sense of Pauli [17, §29]; so were its charge $q$ negligible, the Newtonian equation of motion for a point-like mass (2) would be adequate.
  b) It should be losing the four-momentum through the radiation as effected by the Liénard-Wiechert potentials.

  c) The external force should be small and slowly changing as specified above.
- Inspired by J. J. Thomson, who could made good estimates of both the charge and mass of electron in 1898 by observing its trajectories, we could seek to assess the Dirac, Eliezer and Bhabha kinetic constants of a particular accelerated physical particle by using the differential relation (11) in fitting its velocities. To observe them we could use particle accelerators, since they may accelerate particles of any mass, from electrons and positrons to uranium ions. Such an innovative use of accelerators, for considering also classical kinetic properties of physical particles, might provide new insight into particle acceleration, helpful for improvement of performance and design of accelerator facilities.
- In 1797-98, H. Cavendish performed the first laboratory experiments to measure the force of gravity between masses. This research is ongoing [18], and may inspire the experiments to measure the Dirac, Eliezer and Bhabha kinetic constants of an electrified point-like mass by using the differential relation (11) , e.g. of an electrically charged torque pendulum considered by Saxl [19].

### 6.2. Testing theories

Following Rohrlich [11] and the above preposition about questions we are puting forward

➢ **Proposition about testing theories:** When testing a theory by its implications one should take account of its underlying assumptions.
- The experimentally testable results of theories that take account of RRF are their equations of motion for a classical charged particle. Each equation of motion implies specific asymptotic expansion of velocity analogous to (11), which we can then test experimentally. For instance, up to the order of $\lambda^3$ inclusive, such an asymptotic differential relation corresponding to the Rohrlich equation (9) reads:

$$mc\beta^{(1)} - d(1 - \beta\,\beta\,\cdot\,)\beta^{(2)} \qquad (12)$$
$$+ (d^2/mc)\big[\,\beta^{(3)} - \beta\cdot\beta^{(3)}\,\beta + \beta^{(1)}\cdot\beta^{(1)}\,\beta^{(1)}\,\big] = f\,.$$

Considering the kinetic properties of an electron (or positron), one might check this way Dirac's assumption (7), which he made when deriving Lorentz-Abraham-Dirac equation, i.e. the above hypothesis about the electron.

## 7. Conclusions

A novel interpretation and modification of the Lorentz-Abraham-Dirac equation is put forward that does not ignore the physical premises of the Newtonian mechanics. The asymptotic differential relation (11) for EPM´s velocity provides an improvement on the Lorentz-Abraham-Dirac equation by more precise, non-iterative modeling of RRF´s dynamic effects. It defines new, supplementary kinetic constants of an accelerated electrified point-like mass, which could be assessed by observing its trajectories in an accelerator.

❖ *Any comments, references, suggestions, opinions, and viewpoints will be very much appreciated.*